\documentclass[a4paper,12pt]{article}
\usepackage[utf8]{inputenc}
\usepackage{amsmath}
\usepackage{amssymb}
\usepackage{mathrsfs}
\usepackage{chemarrow}
\usepackage{times}
\usepackage{graphicx}
\begin{document}

\def\MS{{\bf S}}
\def\rd{{\rm d}}
\def\mF{{\bf F}}
\def\mQ{{\bf Q}}
\def\vJ{{\bf J}}
\def\rvB{{\bf B}}
\def\bplus{{\bf +}}
\def\bplus{\mbox{\boldmath$+$}}
\def\ve{{\bf e}}
\def\pp{{\bf p}}
\def\vl{{\bf l}}
\def\vn{{\bf n}}
\def\vv{{\bf v}}
\def\vx{{\bf x}}
\def\vz{{\bf z}}
\def\vnu{\mbox{\boldmath$\nu$}}
\def\vxi{\mbox{\boldmath$\xi$}}
\def\vkappa{\mbox{\boldmath$\kappa$}}


\title{Mesoscopic Kinetic Basis of Macroscopic
Chemical Thermodynamics: A Mathematical Theory
}

\author{Hao Ge$^1$\footnote{haoge@pku.edu.cn} \ and 
Hong Qian$^2$\footnote{hqian@u.washington.edu}\\[6pt]
$^1$Beijing International Center for Mathematical Research (BICMR)\\ Biodynamic Optical Imaging Center (BIOPIC)\\
Peking University, Beijing 100871, P.R.C.\\[1pt]
$^2$Department of Applied Mathematics,
University of Washington\\
Seattle, WA 98195-3925, U.S.A\\[10pt]
}

\maketitle

\begin{abstract}
From a mathematical model that describes a complex
chemical kinetic system of $N$ species and
$M$ elementrary reactions in a rapidly stirred vessel
of size $V$ as a Markov process, we show that a
macroscopic chemical thermodynamics emerges
as $V\rightarrow\infty$.  The theory is applicable to
linear and nonlinear reactions, closed systems
reaching chemical equilibrium, or open, driven systems
approaching to nonequilibrium steady states.  A
generalized mesoscopic free energy gives rise to a
macroscopic chemical energy function $\varphi^{ss}(\vx)$
where $\vx=(x_1,\cdots,x_N)$ are the concentrations
of the $N$ chemical species.  The macroscopic chemical
dynamics $\vx(t)$ satisfies two emergent laws:
(1) $(\rd/\rd t)\varphi^{ss}[\vx(t)]\le 0$; and (2)
$(\rd/\rd t)\varphi^{ss}[\vx(t)]=\text{cmf}(\vx)-\sigma(\vx)$
where entropy production rate $\sigma\ge 0$ represents
the sink for the chemical energy, and chemical motive force
$\text{cmf}\ge 0$ is non-zero if the system is driven under
a sustained nonequilibrium chemostat.   For systems with
detailed balance $\text{cmf}=0$;
and if one assumes the law of mass action,
$\varphi^{ss}(\vx)$ is precisely the Gibbs' function
$\sum_{i=1}^N x_i\big[\mu_i^o+\ln x_i\big]$ for
ideal solutions.  For a class of kinetic systems called
complex balanced, which include
many nonlinear systems as well as many simple
open, driven chemical systems, the
$\varphi^{ss}(\vx)$, with global minimum at $\vx^*$, 
has the generic form
$\sum_{i=1}^N x_i\big[\ln(x_i/x_i^*)-x_i+x_i^*\big]$,
which has been known in chemical kinetic literature. 
Macroscopic emergent ``laws'' are
independent of the details of the underlying kinetics.
This theory provides a concrete example from chemistry
showing how a dynamic macroscopic law can emerge from 
the kinetics at a level below.
\end{abstract}

Ever since the work of J. W. Gibbs and the
influential treatise of Lewis and Randall \cite{lewis-randall-book},
chemical thermodynamics has been one of the most important
theoretical cornerstones of chemical science.  While
temperature is one of the key concepts in the theory,
its origin resides in the mechanical movement of
atoms and molecules, as already clearly articulated by
L. Boltzmann in his mechanical theory of heat \cite{campisi}.
The notion of chemical potential, however, can not be
understood from Newtonian mechanics.  This is best
illustrated through its definition in Gibbs' equation:
\begin{equation}
   \rd U =T \rd S
            -p\rd V
           +\sum_{i=1}^n\rd \mu_iN_i,
\end{equation}
in which
\begin{equation}
 T =  \left(\frac{\partial U}{\partial S}\right)_{V,\{N_i\}},\
 p = -\left(\frac{\partial U}{\partial V}\right)_{S,\{N_i\}},\
\mu_i = \left(\frac{\partial U}{\partial
               N_i}\right)_{S,V,\{N_{j,j\neq i}\}}.
\label{tpmu}
\end{equation}
In graduate texts on Newtonian mechanics, there is a
demonstration of that \cite{gallavotti-book}
$\partial U/\partial S$ is the mean
kinetic energy if one identifies the $S\equiv k_B\ln\Omega$ as the
phase volume of a Hamiltonian dynamics; and
$-\partial U/\partial V$ as the momentum transfer on the
wall of a box that contains gas particles.  While these
demonstrations are not general, they have provided definitive
{\em mechanical} interpretations of the two emergent
thermodynamic quantities.  On the contrary, there is no mechanical
interpretation for the $\partial U/\partial N$  in (\ref{tpmu}).  
It is widely felt that a ``mechanistic'' interpretation of the 
chemical potential $\mu_i$ has to be a
probabilistic one.  We hasten to mention that a living organism
is sustained as a nonequilibrium system neither by a
difference in $T$ nor $p$, rather it is a phenomenon driven
by $\Delta\mu$.  Therefore, a fundamental understanding of
the irreversibility of life requires a deeper understanding of
$\mu$, which is still lacking a rigorous sub-macroscopic
foundation.

{\bf\em The description of the mathematical model.}
We report in this note a recently discovered mathematical
result that provides the chemical potential $\mu$, as a
emergent  macroscopic quantity, an origin in a mesoscopic
description of chemical kinetics.  The mesoscopic description
of a chemical kinetics is based on a system of elementary
chemical reactions with arbitrary complexity; reactions occur
one at a time in a stochastic fashion, as now clearly
demonstrated in single-molecule studies \cite{wemoerner}.
The general setting
has $N$ chemical species and $M$ reactions in a fixed volume
of $V$ \cite{beard-qian-book}:
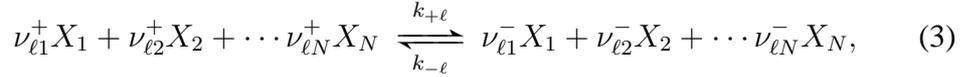
\begin{equation}
     \nu^+_{\ell 1}X_1+\nu^+_{\ell 2}X_2 + \cdots
     \nu^+_{\ell N}X_N  \  \  \underset{k_{-\ell}}{\overset{k_{+\ell}}{\rightleftharpoonsfill{26pt}}}   \  \
      \nu^-_{\ell 1}X_1+\nu^-_{\ell 2}X_2 + \cdots
     \nu^-_{\ell N}X_N,
\label{rxn}
\end{equation}
in which $1\le\ell\le M$.
$\nu_{ij}=(\nu^-_{ij}-\nu^+_{ij})$ are the
{\em stoichiometric coefficients} relating species to
reactions.  In a reaction vessel with rapidly stirred chemical
solutions, the numbers of species $i$ at time $t$ is
denoted by $n_i(t)$.  Our theory assumes that

($i$) Each reaction is microscopically reversible, with
forward rate $r_{+\ell}(\vn(t))$ and backward rate
$r_{-\ell}(\vn(t))$ where $\vn=(n_1,\cdots,n_N)$
denoting the copy numbers of all the species.  Both
$r_{\pm\ell}$ are mathematically non-negative; but their
dependences on $\vn$ are essentailly
arbitrary (with some minor mathematical assumptions).

($ii$)  As an elementrary reaction, each occurs as a Markov
process with exponential waiting time following the
distribution $r(\vn)e^{-r(\vn)t}$.

($iii$) The mesoscopic rate $r_{+\ell}(\vn)$ is the number of
occurrences of the $\ell^{th}$ forward reaction per unit time
in the volume $V$.  Therefore, for a macroscopic system
with extremely large $\vn$ and $V$: $(x_1,\cdots,x_N)\equiv$ $\vx=\vn/V$ are the concentrations, and $R_{\pm\ell}(\vx)=
r_{\pm\ell}(V\vx)/V$ are the concentration-based rates of
the reactions.

($iv$) The chemical reaction system can be either closed or
open.  A closed system has no exchange of matter with
its surrounding; an open system can exchange various
chemical species with its surrounding, which are kept
at constant concentrations.  If all the ``externally buffered''
species are themselves at a chemical equilibrium, the
situation is like a dialysis system, which ultimately reach
a chemical equilibrium both within and with its surrounding.
If, however, there are at least two species that are sustained
at a nonequilibrium condition, then the system eventually
settles in a nonequilibrium state with stationary
concentration fluctuations.  This last scenario is the
biochemical kinetic setup for modeling a living cells
under a continuously chemostat.

	With these rather general assumptions, T. G. Kurtz
has shown in 1972 \cite{kurtz72}
that in the limit of $V\rightarrow\infty$,
the mesoscopic stochastic description of the system of
chemical reactions becomes the following set of rate
equations for the macroscopic kinetics:
\begin{equation}
  \frac{\rd x_i(t)}{\rd t} = \sum_{\ell=1}^M
                \nu_{\ell i}\Big(R_{+\ell}(\vx)-R_{-\ell}(\vx)\Big),
\label{the-ode}
\end{equation}
$1\le i\le N$.   Kurtz's theorem paves the way for a
unified mathematical theory of chemical kinetics in a
rapidly stirred vessel of both small and large size.

	Based on such Markov processes, a mesoscopic statistical
(or stochastic) thermodynamic theory has been developed
in recent years in the field of nonequilibrium statistical
physics.  The most celebrated results from this theory
is the Jarzynski-Crooks equality \cite{jarzynski,crooks}.
The mesoscopic theory
also, for the first time, demonstrated a free energy
blanace equation:  One can introduce a
generalized, nonequilibrium free energy $F^{(meso)}$
for any chemical reaction systems at the mesoscopic level.
Then it can be mathematically shown that this
$F^{(meso)}$ satisfied an instantaneous balance relation
\cite{ge-qian-pre10,esposito-vandenbroeck,qian-jmp}:
\begin{equation}
   \frac{\rd F^{(meso)}}{\rd t} = E_{in} - e_p,
\label{meso-dfdt}
\end{equation}
in which both $E_{in}$ and $e_p$ are non-negative, and
$dF^{(meso)}/dt$ is never positive.
When applied to the chemomechanics of a single ATPase motor
protein, the $E_{in}$ is the amount of chemical energy input
per unit time, e.g., ATP hydrolysis, and $e_p$ is the minimal
amount of heat dissipation \cite{ge-qian-pre13}.
If the motor is actually performing
mechanical work against an external elastic force $f^{ext}$
and moving with velocity $v$, then both $E_{in}$ and
$e_p$ contains the mechanical power $f^{(ext)}v$.

	For a closed chemical reaction system, or an open system
in contact with a single external chemical potential $\mu^{ext}$,
it can be shown that $E_{in}=0$.  In this case,
the $F^{(meso)}$ is indeed the free energy of the molecular 
system \cite{lebowitz-1955,qian-pre01}.
Actually, with the help of detailed balance, the
stochastic Markov theory and Gibbs'
canonical and grand canonical emsemble theories
are proven equivalent.

{\bf\em The kinetic theory of macroscopic
chemical (free) energy.}
We summarize the new mathematical results \cite{part1,part2}.
So far, the equation is only establised in the mesoscopic setting.
So all three non-negative quantities, entropy production
$e_p$, energy input $E_{in}$, and free energy dissipation
$-dF^{(meso)}/dt$ are functions of the volume
parameter $V$.  Furthermore, they are functions of the
probability distribution of the number of each species at time
$t$, $p_V(\vn,t)$, which itself is also a function of $V$.

	Now if we perform the limit of $V\rightarrow\infty$,
Kurtz's theorem tells us that $\vx(t)=\vn(t)/V$ is the solution
to the nonlinear rate equation (\ref{the-ode}).  Furthermore,
the probability theory also suggests that for a wide class of
models, $p_V(V\vx,t)$ can be written as
$\exp\big(-V\varphi(\vx,t)\big)$, and in the stationary state
a $\varphi^{ss}(\vx)$ emerges, which satisfies the following
equation, derived independently by Kurtz in 1978 
\cite{Kurtz1978} and G. Hu in 1986 \cite{hugang}
\begin{equation}
\sum_{\ell=1}^M
                 R_{+\ell}(\vx) \Big[1 - e^{
   \vnu_{\ell}\cdot\nabla_{\vx}\varphi^{ss}(\vx)} \Big]+
    R_{-\ell}(\vx)\Big[1-  e^{ -\vnu_{\ell}\cdot\nabla_{\vx}\varphi^{ss}(\vx)} \Big]
     = 0,
\label{hge}
\end{equation}
where $\vnu=(\nu_1,\cdots,\nu_M)$.  We have
shown that in the ``macroscopic limit'', as intensive
quantities
\begin{subequations}
\begin{equation}
        V^{-1}F^{(meso)} \rightarrow \varphi^{ss}(\vx),
\end{equation}
\begin{equation}
          \frac{\rd\varphi^{ss}(\vx)}{\rd t} =
                  \text{cmf}(\vx)  -\sigma(\vx),
\label{febe}
\end{equation}
\begin{equation}
        V^{-1}e_p \rightarrow  \sigma(\vx)=\sum_{\ell=1}^M
          \Big(R_{+\ell}(\vx)-R_{-\ell}(\vx)\Big)\ln
          \left(\frac{R_{+\ell}(\vx)}{R_{-\ell}(\vx)}\right),
\end{equation}
\begin{equation}
        V^{-1}E_{in} \rightarrow \text{cmf}(\vx)
             = \sum_{\ell=1}^M
          \Big(R_{+\ell}(\vx)-R_{-\ell}(\vx)\Big)\ln
          \left(\frac{R_{+\ell}(\vx)}{R_{-\ell}(\vx)}
            e^{\vnu\cdot\nabla_{\vx}\varphi^{ss}(\vx)}\right).
\end{equation}
\end{subequations}
A macroscopic chemical energy function
$\varphi^{ss}(\vx)$ emerges.  For the
stationary state with large but finite $V$,  
the probability distribution for concentration fluctuations
attains a universal expression
\begin{equation}
     f(\vx|V)=\frac{\Omega(\vx)e^{-V\varphi^{ss}(\vx)}}{
                \Xi(V)}, 
        \text{ where } \
             \Xi(V) =\int \Omega(\vx)e^{-V\varphi^{ss}(\vx)}\rd\vx,
\label{partition-f}
\end{equation}
in which $\Omega(\vx)$ is analogous to the
``degeneracy'' in a partition function calculation.
As a part of the mathematical theory of large deviations in 
a probability distribution, Eq. \ref{partition-f} and alike 
have been considered as the mathematical foundation of 
equilibrium statistical thermodynamics \cite{Hugo09}.   Our
present work is an application of a dynamic version of 
such mathematics.

	What is the mechanistic force corresponding to the
emergent energy function $\varphi^{ss}(\vx)$?
More specifically, how does this ``force'' affect the macroscopic
kinetics?  First, one needs to consciously recognize the vast
separation of time scales in the mesoscopic
and macroscopic kinetics.  That is why the dynamics of
the latter is partially dictated by the stationary behavior of
the former, in the form of $\varphi^{ss}(\vx)$.  It is clear
that $\varphi^{ss}(\vx)$ is a consequence of a global,
infinitely long time behavior of the mesoscopic system.

Eq. \ref{febe} is a macroscopic chemical (free) energy
balance equation, in which an the emergent
{\em chemical motive force} (cmf) characterizes the
force the environment puts upon the kinetic system, and
entropy production rate $\sigma(\vx)$ characterized
the amount of free energy that is dissipated
from the system.

{\bf\em Detailed balance and chemical equilibrium.}
If this global dynamic consequence is in
complete consistency with the local kinetic
$\ln\big[R_{+\ell}(\vx)/R_{-\ell}(\vx)\big]$,
we say the chemical kinetics is in ``equilibrium'':
\begin{equation}
     \ln\left(\frac{R_{+\ell}(\vx)}{R_{-\ell}(\vx)}\right)
  = \exp\Big(-\vnu_{\ell}\cdot\nabla_{\vx}\varphi^{eq}_{\vx}(\vx)\Big),
          \   \forall\vx.
\label{dbc}
\end{equation}
Then $\text{cmf}(\vx)=0$ for all $\vx$, and {\em vice versa}.  
Eq. (\ref{dbc}) formalizes the fundamental insights of G. N. Lewis on
the importantace of detailed balance in chemical kinetics \cite{gnlewis}.
In this case there is {\em a chemical equilibrium between the
local kinetics and its environment that is created by the other
reactions in the same kinetic system.}   If we further
assume the Law of Mass Action, then it can be proven 
mathematically that $\varphi(\vx)$ is actually the
Gibbs function $G(\vx)/k_BT$
$=\sum_{i=1}^N x_i\big(\mu_i^o+k_BT\ln x_i\big)$, further
more Eq. \ref{dbc} is equivalent to $R_{+\ell}(\vx^{eq})=R_{-\ell}(\vx^{eq})$, $\forall\ell$, where $\vx^{eq}$ is the unique
minimum of $G(\vx)$.  Then
Eq. \ref{febe} becomes $\frac{\rd G}{\rd t}=-\sigma(\vx)\le 0$;
and $\vnu_{\ell}\cdot\nabla_{\vx}G(\vx)=\Delta\mu_{\ell}(\vx)$
$=k_BT\ln\big[R_{-\ell}(\vx)/R_{+\ell}(\vx)\big]$,
$\Delta\mu_{\ell}^o=\sum_{i=1}^N \nu_{\ell i}\mu_i^o=k_BT\ln(k_{-\ell}/k_{+\ell})$.

	When Eq. \ref{dbc} is not hold, which is equivalent to
say that $\text{cmf}\neq 0$, then the open, driven
kinetic system eventually settles into a nonequilibrium
steady state $\vx^{ss}$ (or some more complex beheviors
like oscillations,) with $\text{cmf}(\vx^{ss})=\sigma(\vx^{ss})>0$.

{\bf\em Kinetics with complex balance.}
The emergence of $\varphi^{ss}(\vx)$ given in Eq. \ref{hge}
is highly abstract.  One example of this, thanks to the recent
work of Anderson et. al \cite{anderson-2015},
is when the kinetic system
is complex balanced, a notion introduced
by Horn and Jackson in 1972 \cite{horn-jackson}.
This is a class of models which
contains detailed balance, all unimolecular reaction networks,
as well as many open, driven, nonlinear chemical systems.  In
this case, it can be shown that the kinetics equation (\ref{the-ode})
has a unique steady state $\vx^{ss}$ and
\begin{equation}
   \varphi^{ss}(\vx) = \sum_{i=1}^N x_i
          \ln\left(\frac{x_i}{x^{ss}_i}\right) - x_i + x_i^{ss},
\label{shear}
\end{equation}
which is a solution to (\ref{hge}), and the limit of
$-V^{-1}\ln p^{ss}_V(V\vx)$ when $V$ tends infinity.

	We note that for this large class of chemical kinetics,
linear and nonlinear, closed and open, equilibrium and
nonequilibirum, there is a generic, universal expression
for the chemical energy $\varphi^{ss}(\vx)$, Eq. \ref{shear},
which has aleady been in the chemical literature.  This
illustrates the important idea that macroscopic
emergent behavior, such as thermodynamics, should
be independent of the underlying details of the kinetics.

{\bf\em Discussion.}
In summary, macroscopic chemical thermodynamics can
have a rigous mesoscopic, statistical, reaction kinetic
foundation.  Chemical free energy, which is a generalization
of Gibbs' equilibrium free energy, is an emergent quantity
in the macroscopic limit.  It actually has a free energy
balance equation which is different from Newtonian
mechanical energy conservation as well as
Helmholtz-Boltzmann's derivation of the First Law of
Thermodynamics based on their mechanical theory of heat.
This free energy
balance equation is applicable to closed and driven
chemical reaction kinetic systems under isothermal conditions.
	This chemical theory of reaction kinetics also provides
a concrete example for P. W. Anderson's structure of scientific
laws \cite{pwanderson}:
Macroscopic laws are emergent behaviors from the
dynamics of a level below; such laws are insensitive to a large
extend the details of the dynamics.  
It also provides us a tantalizing possibility for high-energy theoretical physics: The universe is made of particles of many kinds and types; the laws that governs their creation, annihilation, transformation, and interactions are the fundamental theory of quantum world.  Yet, at the macroscopic, cosmological scale, one expects emergent laws that govern its macroscopic dynamics in terms of a ``mysterious'' force. Such a force should satisfy an equation like our Eq. \ref{hge}.  Currently, there is an active research program in mathematics that studies this type of nonlinear equations \cite{villani} that includes (\ref{hge}) as well as Ricci flow, the mathematical structure of Einstein's space-time.

Perhaps, chemical science, in addition to providing the living world with tangible materials and useful energy, can also offer some fundamental insights on how the universe works.

\end{document}